# An atomic Boltzmann machine capable of self-adaption

Brian Kiraly[1,†], Elze J. Knol[1,†], Werner M.J. van Weerdenburg[1], Hilbert J. Kappen[2], and Alexander A. Khajetoorians[1,*]

[1]*Institute for Molecules and Materials, Radboud University, Nijmegen, the Netherlands*
[2]*Donders Institute for Neuroscience, Radboud University, Nijmegen, the Netherlands*

[†]These authors contributed equally.

*corresponding author: a.khajetoorians@science.ru.nl

**The quest to implement machine learning algorithms in hardware has focused on combining various materials, each mimicking a computational primitive, to create device functionality. Ultimately, these piecewise approaches limit functionality and efficiency, while complicating scaling and on-chip learning, necessitating new approaches linking physical phenomena to machine learning models. Here, we create an atomic spin system that emulates a Boltzmann machine directly in the orbital dynamics of one well-defined material system. Utilizing the concept of orbital memory based on individual cobalt atoms on black phosphorus, we fabricate the prerequisite tuneable multi-well energy landscape by gating patterned atomic ensembles using scanning tunnelling microscopy. Exploiting the anisotropic behaviour of black phosphorus, we realize plasticity with multi-valued and interlinking synapses that lead to tuneable probability distributions. Furthermore, we observe an autonomous reorganization of the synaptic weights in response to external electrical stimuli, which evolves at a different time scale compared to neural dynamics. This self-adaptive architecture paves the way for autonomous learning directly in atomic-scale machine learning hardware.**

There is a growing pursuit to create brain-inspired devices, or so-called neuromorphic architecture, in order to perform machine learning tasks directly in hardware. These approaches are largely based on using the complex electronic[1-3], magnetic[4-6], and/or optical[7] responses of various solid-state materials to mimic a machine learning model. Most often, only particular aspects of these models can be captured in a given system, thus requiring hybrid schemes that couple different material units, circuitry, and/or external computers in a serial-like fashion to solve a machine learning problem. In such systems, learning is often accomplished by combining the computational primitives of the materials with off-line computers to label data and implement learning algorithms[8-10]. Yet, one of the landmark goals of neuromorphic architecture is to create scalable parallel computing based on autonomous circuits capable of "on-chip" learning[11-13], eliminating the reliance on external computers. To this end, scalable constructions composed of a single material unit will require materials that exhibit dynamical behaviour in both neurons and synapses and self-adapt based on data. Addressing these challenges requires a fundamental understanding of how machine learning functionality, like plasticity, can emerge from the complex dynamics of coupled stochastic ensembles. In this way, new materials, designed for machine learning can drastically improve energy efficiency and computational capabilities; furthermore, these platforms will also provide a vehicle for newly predicted functionality, like quantum machine learning[14-16].

Certain classes of machine learning models, like the Boltzmann machine[17] (BM), are classified as energy-based models, which provide a natural link to physics-based phenomena in materials. The BM, as well as the Hopfield model, are formally equivalent to an interacting Ising spin system where fluctuating spins represent neurons that are linked by memory-bearing weighted interactions that serve as synapses[18]. The realization of the BM relies on the creation of a multimodal energy landscape, strongly linked to the concept of spin glasses[19-21]. Yet, to date there are no well-known spin-based materials that exhibit a tuneable multi-well energy landscape, let alone the full functionality of a BM. Magnetic atoms on surfaces have emerged as a model platform to create tuneable networks of Ising spins[22-25] that can exhibit stochastic dynamics[26,27]. However, the main challenge toward creating magnetic materials that mimic a BM is related to the physical nature of the magnetic exchange interaction: often nearest neighbour interactions dominate, resulting in robust, well-ordered, bistable ground states[26,27] and preventing the formation and adaption of a multi-well potential. Therefore, creating a fully atomic-scale platform in which the spin dynamics of coupled atoms represent a Boltzmann machine requires new fundamental insight into how to realize (1) a multi-well energy landscape that is (2) tuneable via external synapses and (3) inherently self-adaptive based on a suitable learning protocol.

Here, we create a model Boltzmann machine capable of self-adaption realized from the stochastic orbital dynamics of individually coupled Co atoms on the surface of black phosphorus. Taking a bottom-up approach, we start by representing stochastic, binary neurons using the concept of orbital memory. In this representation, the bistable valency states of individual atoms can be locally gated into a stochastic regime by the tip of a scanning tunnelling microscope (STM) (Fig. 1a)[28]. Utilizing atomic fabrication, we effectuate long-range coupling between atoms to produce collective stochastic behaviour, which mimics the steady-state distributions resulting from a multi-well energy landscape for a BM. Additionally, the strongly anisotropic substrate-driven interactions arising from the dielectric behaviour of BP[29,30] allow us to build atomic-scale synapses yielding memory-bearing, multi-valued distributions. Exploiting the separation of time-scales between neural and synaptic dynamics, we demonstrate that the synapses autonomously adapt in response to various input stimuli introduced by the external gating field. This result demonstrates a model material system in which a well-controlled multi-well potential can be tuned to mimic a BM scaled to the atomic limit, where the intrinsic dynamics of the dopants represent a viable route toward atomic-scale autonomous materials for on-chip learning.

***Orbital memory and the Boltzmann machine***

The BM is defined by the following energy equation,

$$E(s|b,w) = -\sum_{i>j} w_{ij} s_i s_j - \sum_i b_i s_i \quad (1)$$

where $s = (s_1,...,s_n)$ represents binary and stochastic spin values and $E(s|b,w)$ defines a steady-state probability distribution for finding the system in a particular state $s$ conditioned on memory bearing and multi-valued weights ($w_{ij}$) and biases ($b_i$) (see methods). A material representation of the BM requires (i) stochastic neural elements $s_i$, and (ii) tuneable and memory-bearing interactions $w_{ij}$, $b_i$ that are ideally self-adaptable in response to external stimuli. It was recently demonstrated that a single Co atom exhibits stable and electrically switchable binary states, based on the concept of orbital memory stemming from its bistable valency[28]. We used individual and identical Co atoms residing on the surface of semiconducting BP (see Supplementary Fig. 1) as building blocks for the BM.

We denote the two orbital memory states of an individual Co atom as spin values $s = 0$ and 1 (Fig. 1a,b). The Co valency is extremely sensitive to its environment: when probed with STM at applied voltages $V_s < V_{th}$ (where $V_{th}$ is the gate voltage threshold required to induce stochastic switching), it exhibits long lifetimes that epitomize its utility as a memory. However, when $V_s > V_{th}$ the atomic valency switches stochastically between its two states (seen in $I_t(t)$ - Fig. 1e with the tip position marked "x" in the STM image in Fig. 1b) with a favourability governed by the local electrostatic environment (see Supplementary Fig. 2c). In this gated regime ($V_s > V_{th}$), we identify the valency $s = 0$ or 1 of an individual Co atom as a neuron in the BM representation (Fig. 1a). In order to correlate the state of $s$ with a discrete and distinct current level measured in the stochastic noise, we reduce the gate voltage below $V_{th}$ to freeze the given state before identifying it with constant-current STM (Fig. 1b and Supplementary Fig. 3). All subsequently displayed STM images were taken in this regime. To clearly distinguish different $s$ configurations, we colour code each distinct level in $I_t(t)$ (Fig. 1e). As the substrate is semiconducting, an applied potential between tip and substrate leads to a local voltage drop within the volume of the semiconductor underneath the tip (shaded area in Fig. 1a), implying the possibility of non-locally gating multiple atoms simultaneously (such as $s_1$ and $s_2$ in Fig. 1a).

A major step toward the realization of an atomic-scale BM was the effectuation of interatomic coupling facilitated by the controlled manipulation of Co atoms with the STM tip (see Supplementary Information for details). When positioning two atoms with interatomic separation less than 4 nm along the BP $x$ direction (armchair = $x$ = [100], zig-zag = $y$ = [010]) (Fig. 1c), we simultaneously gate both atoms with $V_s$ > $V_{th}$, observing the emergence of four distinct and discrete states with appreciable lifetimes in current traces ($I_t(t)$ - constant-height measurement shown in Fig. 1e with the tip position marked "x" in the STM image in Fig. 1c). Using constant-current STM to correlate $I_t(t)$ with the possible state configurations, we map the four possible static $s$ configurations onto each of these current levels (Fig. 1f colour coded in blue with $s$ configurations shown in Supplementary Fig. 3). Integrating over time, we can extract the steady-state probability distribution $\mathcal{P}(s)$ (Fig. 1h) from the measured multi-state current ($I_t(t)$).

The $\mathcal{P}(s)$ distribution shown in Fig. 1i exhibits non-zero populations of all four configurations, indicating that all configurations have appreciable lifetimes. This distribution illustrates the presence of multiple low-lying energy minima, suggesting a multi-well energy landscape of the form (1) with non-uniform couplings and biases. Moreover, the large difference of $\mathcal{P}(s)$ compared to the probability distribution of an isolated atom confirms significant substrate-induced coupling between the atoms (Supplementary Fig. 4). This behaviour scales as we place a third atom in the $x$ direction (Fig. 1d and Supplementary Fig. 5). Above the gate threshold, we observe 7 of 8 possible $s$ configurations (Fig. 1g) and a more complex probability distribution (Fig. 1j). This illustrates appreciable interactions between all atoms beyond nearest neighbour coupling. Additionally, we find that multi-state noise emerges for many different atomic configurations of Co atoms, where the occupational probability of the various *s* configurations strongly depends on the interatomic coupling between individual Co atoms and the gating field (Supplementary Figures 4 and 5). These examples demonstrate that the number of local energy minima increases with atom number. Likewise, this also shows that there is a large phase space of coupling and gating field in which multi-well behaviour emerges, suggesting that precise positioning within a certain range may not be necessary.

*Tailoring synapses using anisotropic coupling*

While atoms coupled in the *x*-direction illustrate a complex probability distribution, the weights $w_{ij}$ that define the probability distribution are fixed by the substrate-mediated interactions. For the BM, it is imperative to create memory-bearing and switchable synaptic weights $w_{ij}$ interlinking the various spins, in order to represent more than one distribution. To do this, we take advantage of the in-plane electronic anisotropy of BP[29,30], which leads to a different coupling between atoms depending on their relative orientation with respect to the surface. In this way, we modify $\mathcal{P}(s)$ of chains of Co atoms formed along the *x*-direction utilizing the memory of a coupled satellite Co atom separated in the *y*-direction (Fig. 2a, c-d). We define the orbital memory states of multiple satellite atoms by $k=(k_1,...k_m)$ (Fig. 2a), quantifying their influence by examining the conditional probability distributions, $\mathcal{P}(s|k)$.

To illustrate this synaptic concept, we show a configuration where a single Co atom (labelled by *k*) is placed approximately 2 nm in the *y*-direction from two coupled atoms, $s_1$ and $s_2$ (Fig. 2c). As seen in Fig. 2e (*k*=0) and Fig. 2f (*k*=1) for the two distinct states of the top Co atom, notable differences can be seen in the multi-state current ($I_t(t)$) when switching is activated. This clearly demonstrates that the *k* atom valency has a substantial impact on the steady-state conditional probability distributions $\mathcal{P}(s|k=0)$ (Fig.

2g) and $\mathcal{P}(s|k=1)$ (Fig. 2h), mimicking an atomic-scale synapse. The distributions $\mathcal{P}(s|k=0)$ and $\mathcal{P}(s|k=1)$ can be modelled as distinct Boltzmann distributions (Eq. 1, 2) using the factor graph representation shown in Fig. 2b. We learn $b,w$ for each value of $k$ by minimizing the Kullback-Leibler (KL) divergence between the empirical probability distribution $\mathcal{P}(s|k)$ of the states and the BM distribution $\mathcal{P}(s|b,w)$. The results of this fitting for both $k=0$ and $k=1$ are shown in Table 1. The conditions $k = 0$ or 1 result in significant modifications to the synaptic coupling ($w_{12}$) and the biases ($b_1,b_2$), directly influencing the BM energy landscape (1). Thus, the valency of the satellite Co atom acts as discrete parametrization for $b,w$. Therefore, this three-atom representation can be mapped to a BM (Fig. 2b, see also Supplementary Fig. 7) which contains two neurons and one binary synapse.

In order to be able to represent and learn a larger set of distributions, we introduce four $k$ atoms coupled to three $s$ atoms (Fig. 3a). With two states associated with each $k$ atom, there are 16 possible $k$ configurations, so that we can represent 16 possible distributions $\mathcal{P}(s|k)$ over the spins $s$. We show a subset of five possible $k$ configurations in Fig. 3. Figure 3a shows constant-current STM images, below the gate threshold, identifying the valency of each $k$ atom for the five illustrated configurations. We employ the same characterization as in Fig. 2, measuring $I_t(t)$ above the gate threshold (Fig. 3b) and extracting $\mathcal{P}(s|k)$, for each value of $k$. These five distributions are plotted in Fig. 3c, illustrating highly varying changes in the distributions, depending on $k$. For nearly every studied $k$-configuration, we observe significantly different distributions for $\mathcal{P}(s)$, as shown by analysing the KL divergence between all $\mathcal{P}(s|k_i)$ distributions (Supplementary Fig. 11, see also the corresponding analysis for Fig. 4 in Supplementary Fig. 12). Using the same modelling described above, for three spins with individual biases $b$, pairwise couplings $w$, and a three spin coupling $w_{123}$, we find different solutions $\mathcal{P}(s|b,w,w_{123}) = \mathcal{P}(s|k)$ for different values of $k$, indicating that different $k$ configurations correspond to different synapse and

bias values. As seen in Table 2 (also Supplementary Tables 1-3), the mapping of **k** onto **b,w,**$w_{123}$ is highly non-linear, likely related to the complex electrostatic environment of the atomic ensemble.

*Separation of time scales and self-adaption*

In Fig. 2 and 3, we showed that each **k** configuration results in a different distribution $\mathcal{P}(s|k)$ of the **s** configurations, and thus emulates synapses and biases of the BM, scaled to the atomic limit. Hence, changes in **k** result in changes to the BM parameters (Table 1-2) and thus could be used to implement a learning rule in a regime where the **k**-state lifetime is nearly infinite. One possibility to input complex signals onto this system would be to impose multiplexed AC signals on the DC gate and implement an external computer to modify the values of **k** according to an alternative training algorithm, which still needs to be designed, specifically for discrete weights[31]. As a step toward autonomous behaviour, it has been shown that spike-timing dependent neural dynamics can be coupled to synapses, and as such, these synapses can evolve in response to neural stimulation[12,13]. Indeed, in biological systems, learning is a dynamical process identified with synaptic evolution, namely a distinct time scale in which synapses evolve based on exposure to stimuli. In this way, while neurons change their state on the order of milliseconds, learning, e.g. in the context of long-term potentiation, occurs over minutes to hours[32]. Therefore, it is essential to explore materials capable of hosting both neural and synaptic dynamics, that exhibit plasticity, and subsequently exploring new learning models that exploit their coupled network dynamics.

As a step toward this end, we subsequently explore the separation of time scales in our system by quantifying the dynamics of the **k**-configurations and their correlation with the steady-state neural distributions and environmental stimuli. We studied the **k** dynamics in a seven atom BM (shown experimentally in Fig. 4a and illustrated schematically in Fig 4b) over sufficiently long time scales (300-2000 minutes per distribution) in order to allow for the spontaneous evolution of $k$. We probed the response of $\mathcal{P}(k)$ to changes in the environmental stimuli in the form of small changes in an applied offset voltage ($\varepsilon = V_{off} = V_s - V_{th}$, where $V_{th} \approx 400$ mV), in a regime where the total voltage is far above the stochastic switching threshold. We performed the measurements by initializing the ensemble into a random $s$ and $k$ configuration before measuring $I_t(t)$ (Fig. 4c) at a specified $V_{off}$. At each environmental condition, measurements were conducted until the $\mathcal{P}(s,k)$ converges to a representative and distinct steady-state distribution (Fig. 4d-f, see also Supplementary Fig. 9 and 10). Generally, the time scale for convergence of $\mathcal{P}(k|\varepsilon)$ for given $\varepsilon$ is 1000-4000 times longer than for $\mathcal{P}(s|k)$ for given $k$. As seen in Fig. 4d-f, $\mathcal{P}(k|\varepsilon)$ reaches a steady-state that is extremely sensitive to changes in the stimulus on the order of 10 mV. When $V_{off} = 200$ mV, $k = (0000)$ is the most favourable synaptic configuration ($\mathcal{P}(k=0000|\varepsilon_3) = 0.43$), while at $V_{off} = 160$ mV the favourability moves to $k = (0100)$ ($\mathcal{P}(0100|\varepsilon_1) = 0.38$). In other words, $\mathcal{P}(k|\varepsilon)$ autonomously adapts in response to small variations in the gate voltage. In order to demonstrate that this response is not random, we did the following control experiment; we measured at (i) $\varepsilon_1$, (ii) $\varepsilon_3$, and again at (i) $\varepsilon_1$, while checking the conditional probability $\mathcal{P}(k|\varepsilon)$. The measurements confirm that the $\mathcal{P}(k|\varepsilon_1)$ distribution remains the same before and after the intermediary stimulus $\varepsilon_3$. Furthermore, it is clear that the evolution of $\mathcal{P}(k|\varepsilon)$ is non-linear in $V_{off}$; this is nicely exemplified when examining the $\varepsilon$-dependent probability for $k = (1000)$: $\mathcal{P}(k=1000|\varepsilon_1) = 0.09$, $\mathcal{P}(k=1000|\varepsilon_2) = 0.33$, $\mathcal{P}(k=1000|\varepsilon_3) = 0.20$, which is maximum at $V_{off} = 180$ mV ($\varepsilon_2$). Visualizing such non-linearity is aided by considering the additional complexity in the joint probability distribution $\mathcal{P}(s,k|\varepsilon) = \mathcal{P}(s|k,\varepsilon)\mathcal{P}(k|\varepsilon)$ (shown in blue in Fig. 4d-f). What is clearly evidenced in Fig. 4d-f, is that distinct steady-state $\mathcal{P}(k|\varepsilon)$ distributions are correlated with distinct environmental stimuli (also seen in Supplementary Fig. 8). In other words, the $k$ configurations self-adapt over a longer time scale, conditioning their state on the input stimulus according to a multi-modal energy landscape defined by the stimuli, analogous to the landscapes of the spins. While we only characterized the BM response to various DC stimuli, we observed a strong non-linear frequency and amplitude dependence in the single neurons' $\mathcal{P}(s)$ response (Supplementary Fig. 2). This suggests that the neural and synaptic dynamics exhibit a complex, coupled, and rich landscape. As such, multiplexing may potentially be used to encode complex signals in the spectral components of an AC signal[4] and introduced via the tunnelling barrier, if a sufficient learning algorithm is developed.

In contrast to hybrid approaches, here both neurons and synapses are contained in a single material and the physics of the coupling separates the time scales of the *s* and *k* dynamics by multiple orders of magnitude. In this situation, we equate neural computation with the *s* variables which exhibit fast dynamics and learning with the slower environmentally dependent evolution of the synaptic distribution ($\mathcal{P}(k|\varepsilon)$). In this picture, the neural and synaptic dynamics are coupled and therefore standard learning algorithms cannot be used. When the time scales are well separated the neural activity equilibrates to a stationary distribution and one may be able to develop a learning algorithm that acts on this statistical distribution of neural activity[33]. Note that learning in this context implies non-trivial, coupled changes to both the synaptic and neural distributions. This is different from current hybrid approaches for on-chip learning where, for example, synaptic strength is conditioned by neural inputs[11-13].

It remains to be seen if the complex dynamics demonstrated here can be scaled up to a larger number of atoms. The observation of multi-state noise rules out a description dominated by nearest neighbour interactions and suggests the presence of a complex, spatially varying and time dependent mean field (Supplementary Fig. 2, 4, 5, and 6). In order to better understand the potential for scaling, it is essential to quantify the interplay between the substrate-mediated interactions, the role of dielectric screening and the external gating field, in order to clarify how each contributes to machine learning functionality. To probe this experimentally, larger scale cobalt ensembles need to be created, which may be possible with electronic mediated growth mechanisms[29]. It remains, however, an open challenge to fabricate precise atomic-scale gate structures, as is typically done for dopants in silicon based devices[34], to gate and read-out the BM state. Further work is needed to explore new material systems which might exhibit orbital memory behaviour[35] up to higher operating temperatures.

*Conclusion*

In conclusion, we have constructed a network of stochastic orbital memories derived from Co atoms on BP that emulates the behaviour of a Boltzmann machine scaled to the limit of individual atoms. When controllably coupling and gating individual Co atoms, we observed the onset of a tuneable multi-modal landscape that is largely unexpected in model Ising materials. Utilizing the anisotropic behaviour of black phosphorus, we were able to create atomic-scale synapses that tuned and stored the weights/biases. We have further shown a separation of neural and synaptic time scales, larger than three orders of magnitude. This allowed us to confirm that the synaptic, memory-bearing atoms autonomously reorganize in response to an input DC stimulus. In addition to this self-adaption, we showed that the neural dynamics exhibits rich and complex phase spaces in response to both DC and AC voltage signals. Using orbital memory is potentially much more energy efficient than approaches which rely on phase changes in materials (see Supplementary Information). Furthermore, future developments of learning algorithms that consider the coupled dynamics of both neurons and synapses at their different time scales will allow this system to physically compute using the spins in materials, drastically reducing the computational costs associated with Monte Carlo sampling[6,36].

**Acknowledgements**

This project has received funding from the European Research Council (ERC) under the European Union's Horizon 2020 research and innovation programme (grant agreement No 818399). This research was funded in part by ONR Grant N00014-17-1-2569. A.A.K. and E.J.K. acknowledge the NWO-VIDI project "Manipulating the interplay between superconductivity and chiral magnetism at the single-atom level" with project number 680-47-534. B.K. acknowledges NWO-VENI project "Controlling magnetism of single atoms on black phosphorus" with project number 016.Veni.192.168.

**Author Contributions**

B.K. and E.J.K. performed the experiments under the direction and supervision of A.A.K. B.K. and E.J.K. developed the data analysis, while B.K, E.J.K, H.J.K and A.A.K participated in the scientific analysis. W.M.J. van W. developed the AC experimental setup. H.J.K. performed the Boltzmann machine modelling. A.A.K. and H.J.K. designed the experiments. The manuscript was written by B.K., E.J.K, H.J.K and A.A.K.

**Competing Interests**

The authors declare no competing interests.

**Materials and Correspondence**

Correspondence to Alexander Khajetoorians.

**Data Availability**

The data from this work can be obtained from the corresponding author upon reasonable request.

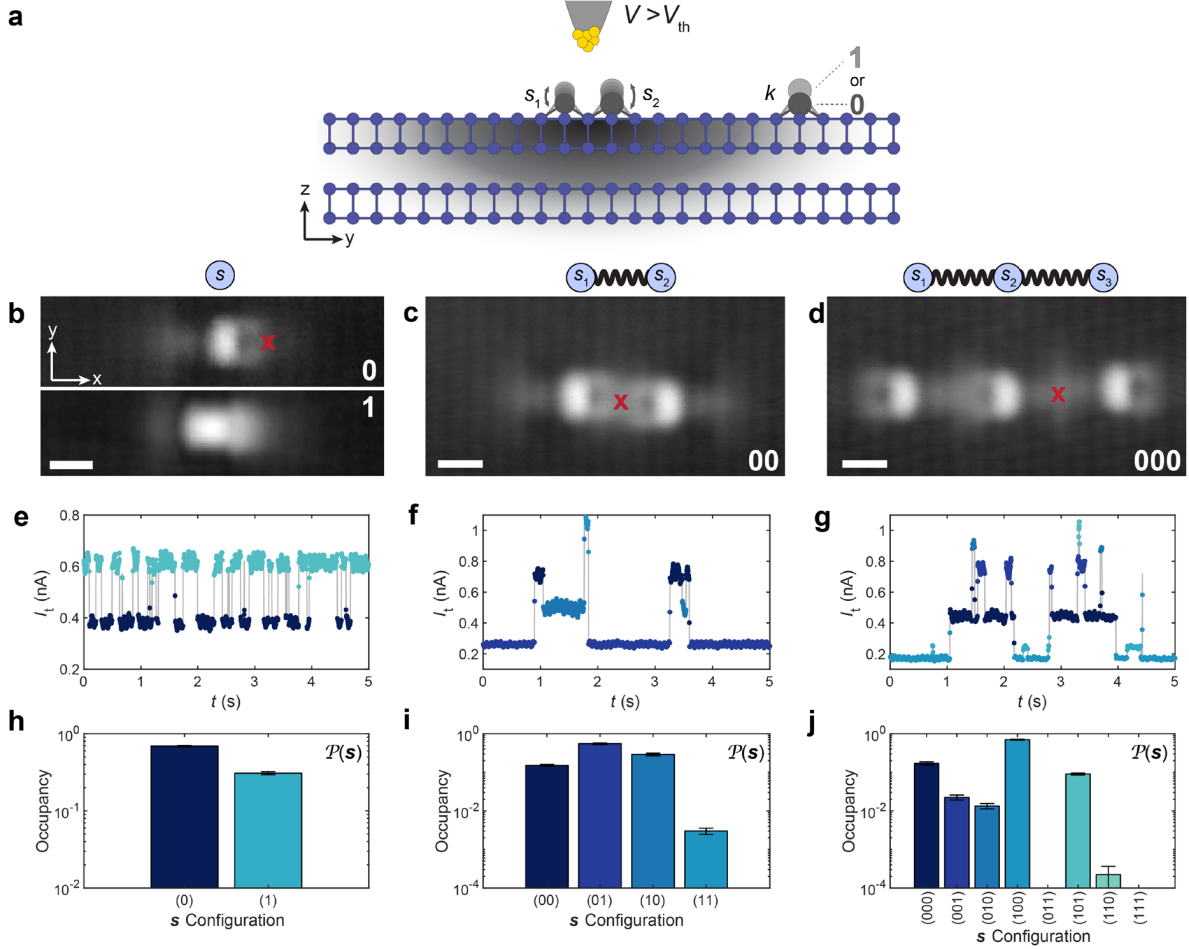

**Figure 1. Neural dynamics from coupled cobalt atoms on black phosphorus**. **a**, Schematic illustrating experimental setup with STM tip gating the system above the switching threshold. Here, due to differences in local band bending (grey) $s_1$ and $s_2$ atoms switch stochastically, while $k$ remains static. **b**, Isolated cobalt atom in two valency configurations ($s = 0$ and 1) with armchair = $x$ = [100] and zig-zag = $y$ = [010] crystallographic directions defined ($V_s$ = -400 mV, scale bar = 1 nm). **c**, Two cobalt atoms in state $(s_1, s_2) = (0,0)$ and **d**, three cobalt atoms in state $(s_1, s_2, s_3) = (0,0,0)$ separated by approximately 5-6 unit cells along the armchair direction ($V_s$ = -400 mV, scale bar = 1 nm). **e**, Two-state current signal $I_t(t)$ observed in constant-height when gating the single cobalt atom at the red "x" in a with $V_s$ = 550 mV. **f-g**, Multi-state current signal $I_t(t)$ measured at the red "x" in **c** and **d** with $V_s$ = 550 mV. **h**, Time-integrated probability distribution ($\mathcal{P}(s)$) for each valency ($s$ = 0 or 1) in the isolated Co atom. **i-j**, $\mathcal{P}(s)$ for dimer and trimer systems, respectively. Error bars show 95% confidence intervals.

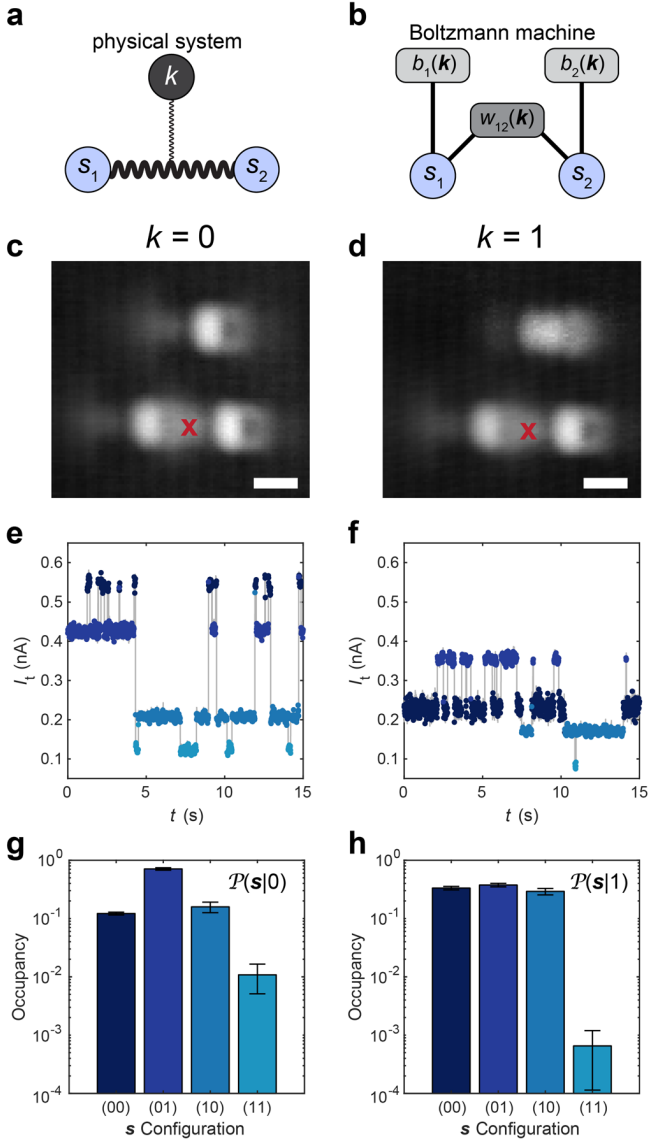

**Figure 2. Construction of a binary atomic synapse via anisotropic coupling. a**, Schematic representation of the three atom system in **c** and **d**, where the two atoms $s_1$ and $s_2$ are strongly coupled and the $k$ atom is weakly coupled to $s_1$ and $s_2$. **b**, Factor-graph representation of the Boltzmann machine mapped to the experimental configuration depicted in **a** and shown in **c** and **d**. **c-d,** Constant-current STM topography with $k=0$ and $k=1$, respectively ($V_s$ = -400 mV, scale bar = 1 nm). **e-f,** $I_t(t)$ measured at the red "x" in **c** and **d** with $V_s$ = 500 mV. **g-h,** Conditional, time-integrated probability distributions $\mathcal{P}(s|k)$ for $k=0$ and $k=1$, respectively. Error bars show 95% confidence intervals.

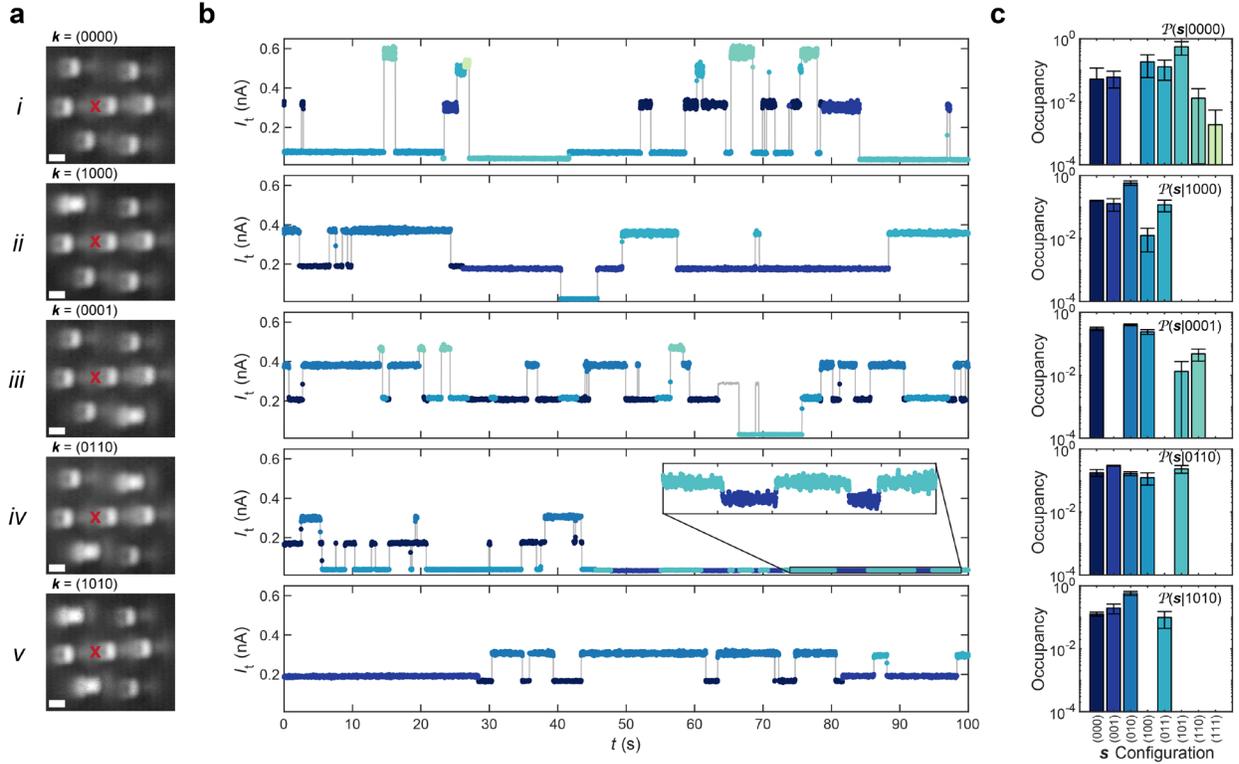

**Figure 3. Multi-valued synapses. a,** Constant-current STM images of a seven atom cobalt ensemble with three s atoms and four k atoms ($V_s$ = -400 mV, scale bar = 1nm). The **k** configurations are: *i* = **k** = (0000), *ii* = **k** = (1000), *iii* = **k** = (0001), *iv* = **k** = (0110), and *v* = **k** = (1010). **b,** $I_t(t)$ measured at the red "x" positions in **a** with $V_s$ = 500 mV. **c,** Conditional probability distributions, $\mathcal{P}(s|k)$. Error bars show 80% confidence intervals.

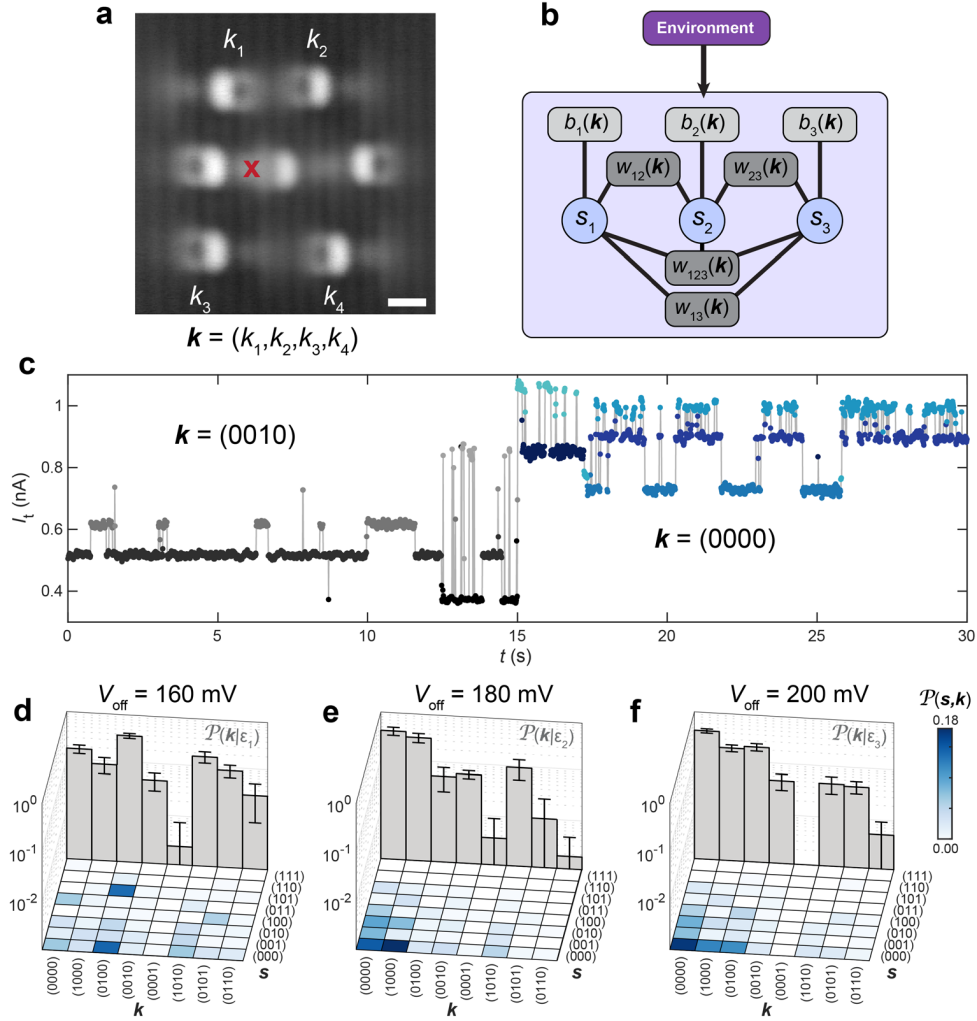

**Figure 4. Synaptic dynamics and self-adaption. a,** Constant-current STM image of a seven atom cobalt ensemble with three *s* atoms and four *k* atoms ($V_s$ = -400 mV, scale bar = 1nm). **b,** Factor graph representation of the Boltzmann machine used to model the atomic ensemble in **a**, with a diagram illustrating the influence of the environment on the BM. **c,** $I_t(t)$ taken at the red "x" in **a** with $V_{off}$ = 160 mV showing a spontaneous modification of the system's *k* (before in grey and after in blue), which is observed through the distinct current levels and stochastic dynamics. **d-f,** Probability distributions $\mathcal{P}(s,k|\varepsilon)$ shown in blues and *s*-integrated probability distribution ($\mathcal{P}(k|\varepsilon)$ – grey histograms), given the environmental conditions $\varepsilon_1 \equiv V_{off}$ = 160 mV, $\varepsilon_2 \equiv V_{off}$ = 180 mV, $\varepsilon_3 \equiv V_{off}$ = 200 mV. Error bars show 95% confidence intervals.

|     | $k = 0$ | $k = 1$ |
| --- | --- | --- |
| $w_{12}$ | -4.48 | -6.2 |
| $b_1$ | 0.28 | -0.14 |
| $b_2$ | 1.78 | 0.1 |

**Table 1. Weights and Biases for the *k* configurations shown in Fig. 2.**

|     | $k$=(0000) | $k$=(1000) | $k$=(0001) | $k$=(0110) | $k$=(1010) |
| --- | --- | --- | --- | --- | --- |
| $w_{12}$ | 18.85 | -19.94 | 3.19 | -21.41 | -2.62 |
| $w_{13}$ | 9.39 | -18.43 | 17.34 | -7.08 | -1.53 |
| $w_{23}$ | 22.25 | -8.06 | 4.79 | -22.28 | -3.31 |
| $w_{123}$ | -16.87 | 13.35 | -10.15 | 14.43 | 2.20 |
| $b_1$ | -11.39 | 7.47 | -7.84 | 10.45 | -12.86 |
| $b_2$ | -23.13 | 12.66 | -5.66 | 16.30 | 3.06 |
| $b_3$ | -14.16 | 11.52 | -16.91 | 13.00 | 2.25 |

**Table 2. Weights and Biases for the *k* configurations shown in Fig. 3.**

**Methods**

*Scanning Tunnelling Microscopy*

STM measurements were performed under ultrahigh vacuum ($< 1\mathrm{x}10^{-10}$ mbar) conditions with an Omicron low-temperature STM at a base temperature of 4.4 K, with the voltage applied to the sample. The typical time resolution of these experiments was approximately 1 ms. All STM images were acquired by means of constant-current feedback. All $I_t(t)$ measurements were acquired with the tip at a constant height and the feedback loop turned off. For all measurements in the main text, the tip height was stabilized with constant-current feedback on the bare black phosphorus, at $I_t = 20$ pA, $V_s = -400$ mV.

Electrochemically etched W tips were used for measurements; each tip was treated in situ by electron bombardment, field emission, as well as dipped and characterized on a clean Au(111) surface. Black phosphorus crystals were purchased from HQ graphene and subsequently stored in vacuum (< 1x10$^{-8}$ mbar). The crystals were cleaved under ultrahigh vacuum conditions at pressures below 2x10$^{-10}$ mbar, and immediately transferred to the microscope for in-situ characterization. Cobalt was evaporated directly into the STM chamber with $T_{STM}$ < 5 K for the entire duration of the dosing procedure. Atomic manipulation of the cobalt atoms was done by dragging the atoms in constant-current feedback mode with -130 mV < $V_s$ < -100 mV and 6 nA < $I_t$ < 12 nA (see Supplementary Information).

*Computing probability distributions*

To identify each *s* switching event in the $I_t(t)$ data, an algorithm was used that determines the locations of abrupt changes (steps) in data. The algorithm split the data into segments for which the difference between the residual error and the mean of the data in the segment is minimal[37,38]. Where necessary, the data was pre-processed by smoothing and corrected for the z-drift between tip and sample. After identifying all the switches in the data, the (local) mean of all the segments ($I_{mean,i}$) was calculated between switching events. Discrete maxima in the histogram of the mean and/or raw current signal were used to identify a target current ($I_{s,i}$) for each *s* configuration; individual segments in $I_t(t)$ were assigned to *s* configurations based on the smallest value for |$I_{mean,i}$ - $I_{s,i}$|. To compute $\mathcal{P}(s)$, the histogram of the discretized data was computed and the total values for each *s* configuration were normalized to the total length of the measurement. To acquire distributions $\mathcal{P}(k)$, each *k* switch was identified manually. For both $\mathcal{P}(s)$ and $\mathcal{P}(k)$, all data is shown after acquiring at minimum two times longer than the convergence time.

# An atomic Boltzmann machine capable of self-adaption

Brian Kiraly[1,†], Elze J. Knol[1,†], Werner M.J. van Weerdenburg[1], Hilbert J. Kappen[2], and Alexander A. Khajetoorians[1,*]

[1]*Institute for Molecules and Materials, Radboud University, Nijmegen, the Netherlands*

[2]*Donders Institute for Neuroscience, Radboud University, Nijmegen, the Netherlands*

[†]These authors contributed equally.

*corresponding author: a.khajetoorians@science.ru.nl

**Contents**

# Tip preparation and characterization

The tip was first prepared on a Au(111) substrate by dipping and pulsing until the tip was stable, sharp, and displayed a d$I$/d$V$ spectrum characteristic of the Au(111) surface state. Black phosphorus (BP) samples were then cleaved in ultra-high vacuum (UHV) and inserted immediately into the scanning tunnelling microscope (STM). Clean BP surfaces are predominantly featureless and exhibited $p$-type doping with a band gap of approximately 0.30 eV - 0.33 eV, as reported previously[1,2] (similar to Supplementary Fig. 1b).

A large-scale image showing the BP after deposition of Co at 4.6 K (see methods for details), is given in Supplementary Fig. 1a. As seen in Supplementary Fig. 1a and described previously[3], Co atoms are found in both the hollow binding site and the top binding site after deposition, with a higher probability of residing in the top site. In order to determine the work function of the tip, we also characterized the position of the charging feature in the d$I$/d$V$ spectrum[3] for a Co atom in the hollow site and state 0 (Supplementary Fig. 1c). In this work, the condition of the tips was primarily such that the position of the charging feature was found between 280 mV and 400 mV. The energetic position of the charging feature also corresponds roughly to the onset of stochastic switching between the two Co valencies (denoted 0 and 1 in this work) in the hollow binding site. As such, the position of the charging feature can be used to characterize the offset voltage, $V_{off}$, above which stochastic switching is observed. In multi-atom ensembles, however, the picture is a bit more complicated. The charging energy is no longer well-defined (can vary between atoms), thus the offset voltage can only be approximated by the charging-related feature.

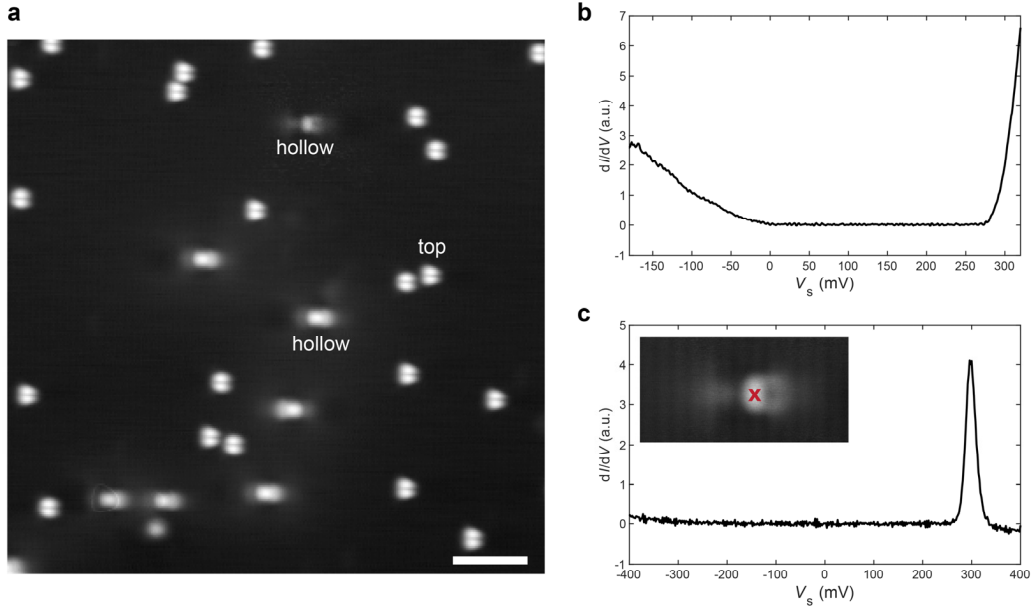

**Supplementary Figure 1. Overview image and spectroscopy of individual Co atoms on BP. a,** Characteristic overview image of Co atoms on BP after deposition at 4.6 K ($V_s$ = -400 mV, $I_t$ = 20 pA, scale bar = 5 nm). As labelled in the image, the Co atoms are found in both top and hollow binding sites directly after deposition. **b,** d$I$/d$V$ spectrum on the bare BP surface, illustrating the characteristic semiconducting band gap. **c,** d$I$/d$V$ spectrum taken at the red "X" on the cobalt atom in the hollow binding site and the 1 state (see Fig. 1). The characteristic charging peak, seen here at approximately 300 mV, was used for tip characterization.

## Atomic manipulation

As seen in Supplementary Fig. 1a (large scale image) there are multiple species of Co atoms on the surface, residing on different adsorption sites (top/hollow). The top-site atoms can be irreversibly moved into the hollow site by sweeping the applied sample bias to $|V_s| > 700$ mV with the tip positioned near the centre of the atom. The most reliable atomic manipulation was performed on the top-site species, which were dragged or pushed along the BP zigzag rows ($y$ direction). Typical parameters for atomic manipulation of top-site Co atoms are: 6 nA < $I_t$ < 12 nA and -130 < $V_s$ < -100 mV. After reaching the desired location, the top-site atoms are moved into the hollow site. All measurements are performed on ensembles of hollow-site atoms.

## Single atom switching characteristics

### DC response

The switching characteristics of individual Co atoms have been studied in detail previously[3]. Of particular interest here, are the characteristics of the single atom response with respect to the applied

DC sample voltages. In order to characterize the steady-state probability distribution for a single atom, here we define the asymmetry as $A = \mathcal{P}(0) - \mathcal{P}(1)$. Consistent with previous results[3], we observe that $A$ is directly proportional to $V_s$ (Supplementary Fig. 2c). We note that this behaviour is quite sensitive to the work function of the STM tip as seen by the shifts in the $A$ vs. $V_s$ data shown in Supplementary Fig. 2c for different tip terminations or microtips. The direct proportionality between $A$ and $V_s$ was, however, always present with tips exhibiting characteristic charging features between 280 mV and 400 mV as described above. Therefore, to first approximation, the relative value of $V_s$ between different tips is not always equal, resulting from variations in tip-induced band bending.

AC response

To investigate whether individual Co atoms exhibit a sensitivity to time-dependent stimuli, we studied the effects of both sinusoidal/triangular AC voltages, added onto $V_s$, on the probability distribution $\mathcal{P}(s)$ of individual atoms. An active adder was used to add the AC signal from an arbitrary waveform generator ($V_{AC}$) to the DC bias ($V_{DC}$) from the control electronics, and we confirmed the fidelity of the ingoing waveform at various amplitudes and frequencies. $I_t(t)$ traces were taken with a constant $V_{DC}$ below the switching threshold here ($V_{off} > V_{DC} = 350$ mV) and an AC voltage was added with peak-to-peak amplitude $V_{AC} = 300$ mV. The cumulative DC + AC voltage periodically exceeded the gate threshold at intervals determined by the waveform (sine/triangle) and frequency. To avoid small changes in individual $I_t(t)$ traces, we started the measurements at the same $x, y, z$ tip position: the tip was stabilized on the highest point of the single atom in the $s = 0$ state, found by atom tracking. Thereafter the tip was moved in constant height to the measurement position, 1 nm off the highest point of the atom.

As a single Co atom switches between two states, $s = 0$ and $s = 1$, the probability distributions $\mathcal{P}(\mathbf{s})$ for different frequencies and waveforms are easy to compare by computing $A$. As seen in Supplementary Fig. 2a-b, the asymmetry exhibits strongly non-linear behaviour below approximately 200 Hz. Furthermore, when comparing measurements with the same tip and the same atom, yet different applied waveforms (pairs of cyan or black points in Supplementary Fig. 2a-b), it is clear that while the overall trend of increasing $A$ with decreasing $f$ remains, the details of each measurement are quite different. Considering that the triangle wave is comprised of sine waves with the fundamental frequency and higher ($f_n = 3f_0$) harmonics, the differences in these responses indicates that the single neuron is sensitive to multiple frequency inputs. This sensitivity is significant, in that it indicates that frequency-encoded signals can be distinguished by single neurons.

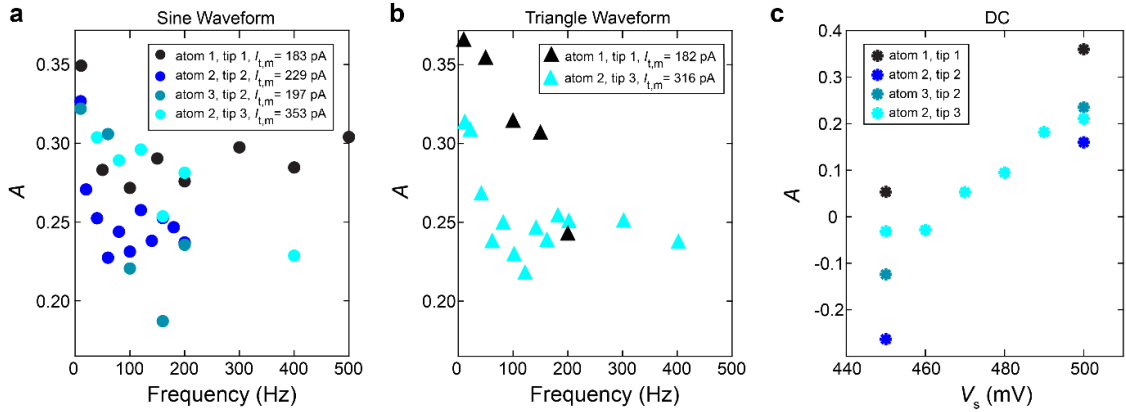

**Supplementary Figure 2. Frequency dependent response of a single cobalt atom on BP. a-b,** Asymmetry of the probability distribution $\mathcal{P}(s)$ as a function of frequency of the applied AC voltage for sine and triangle waveforms, respectively. For the triangle waveform, the plotted frequency is the fundamental or lowest frequency. The experiment was performed on different atoms with different microtips (see legends). For each plotted data point multiple $I_t(t)$ traces were taken. Small tip variations led to differences in the mean current $I_{t,m}$ for each $I_t(t)$ measurement (see legends). Individual current traces with a mean current within 14 pA or roughly twice the standard deviation of the overall mean $I_{t,m}$ were used to calculate $\mathcal{P}(s)$, $A$. **c,** Asymmetry of the probability distribution $\mathcal{P}(s)$ as a function of $V_s$, for the different microtips that were used in the AC experiment in **a** and **b**. For all microtips used in this experiment, the asymmetry increases with increasing $V_s$.

## Detailed characterization of coupled *s*-atoms

### Identification of valency configurations

In order to correlate the atomic valency to a specific level in the current signal $I_t(t)$ the following procedure was used. Stochastic current noise was measured by applying a voltage with $V_s > V_{th}$ for a specified period of time, in which the mean lifetimes is significantly longer than 10 ms. Immediately after this period of time, the applied voltage was within 10 ms reduced to $V_s < V_{th}$ in order to freeze the final atomic valency. The state of the atom or atoms (*s*) could then be measured using constant-current STM with $V_s < V_{th}$ (see Supplementary Fig. 3). This procedure was repeated 2-6 times for each state in order to ensure no additional switches occurred during the voltage ramp to $V_s < V_{th}$. For extremely short-lived states, the $I_t(t)$ curve was manually terminated after the system switched into the target current level. The subsequent identification procedure was the same. Using this procedure, STM topography images and atomic configurations of the different valencies from Fig. 1 are shown in Supplementary Fig. 3.

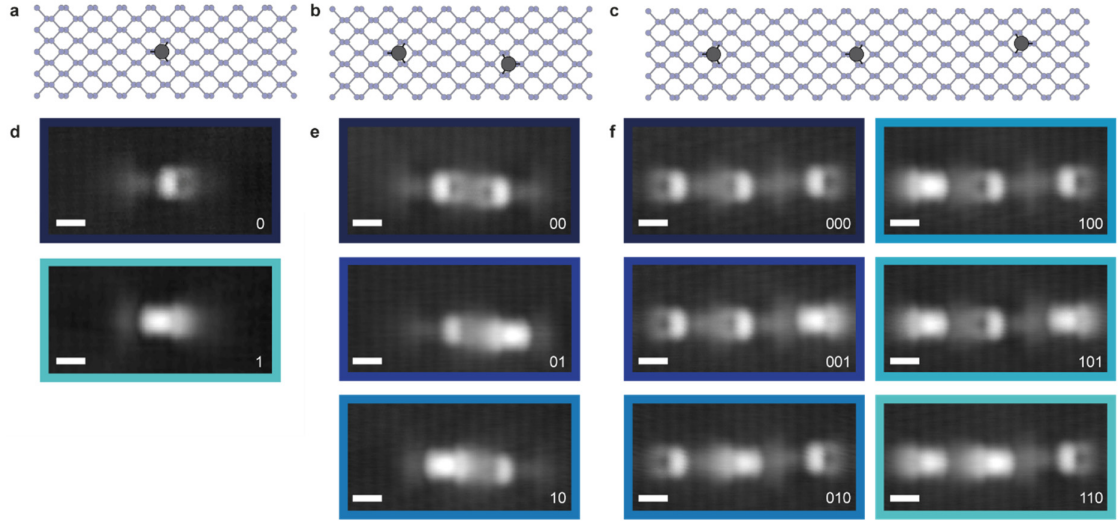

**Supplementary Figure 3. Schematics and valency configurations of coupled Co atoms on BP. a-c,** Schematic of the BP lattice structure with the atomic positions of the coupled Co atoms indicated. **d-f,** STM topographies of the different configurations that were observed in the $I_t(t)$ traces of the configurations seen in Figure 1 ($V_s$ = -400 mV). The colours of the boxes around the images correspond to the current levels in Figure 1. For the case of two and three coupled atoms **e-f**, the (11) and (111) configurations could not be trapped due to their short lifetime.

Influence of interatomic separation on neural dynamics

The multi-well behaviour for coupled Co atoms in the $x$ direction ($s$ atoms) persists for interatomic separations between roughly 1.6 nm and 3.5 nm (four to eight BP lattice sites). Five different configurations of two coupled atoms are shown in Supplementary Fig. 4. As a single Co atom on BP can have two different orientations depending on the symmetry of the hollow binding site (for instance, Supplementary Fig. 4a,b), in addition to the two different orbital configurations ($s$ = 0/1); indeed both the orientation and separation of the Co atoms influence the underlying coupling, as seen in Supplementary Fig. 4. The multi-well potential is present in all configurations shown in Supplementary Fig. 4 and later in Supplementary Fig. 5, as evidenced by the robustness of the multi-state noise to various separations. The exact $\mathcal{P}(\mathbf{s})$ distribution, however, is different for all configurations, indicating that the quantitative behaviour is quite sensitive to the precise atomic positions.

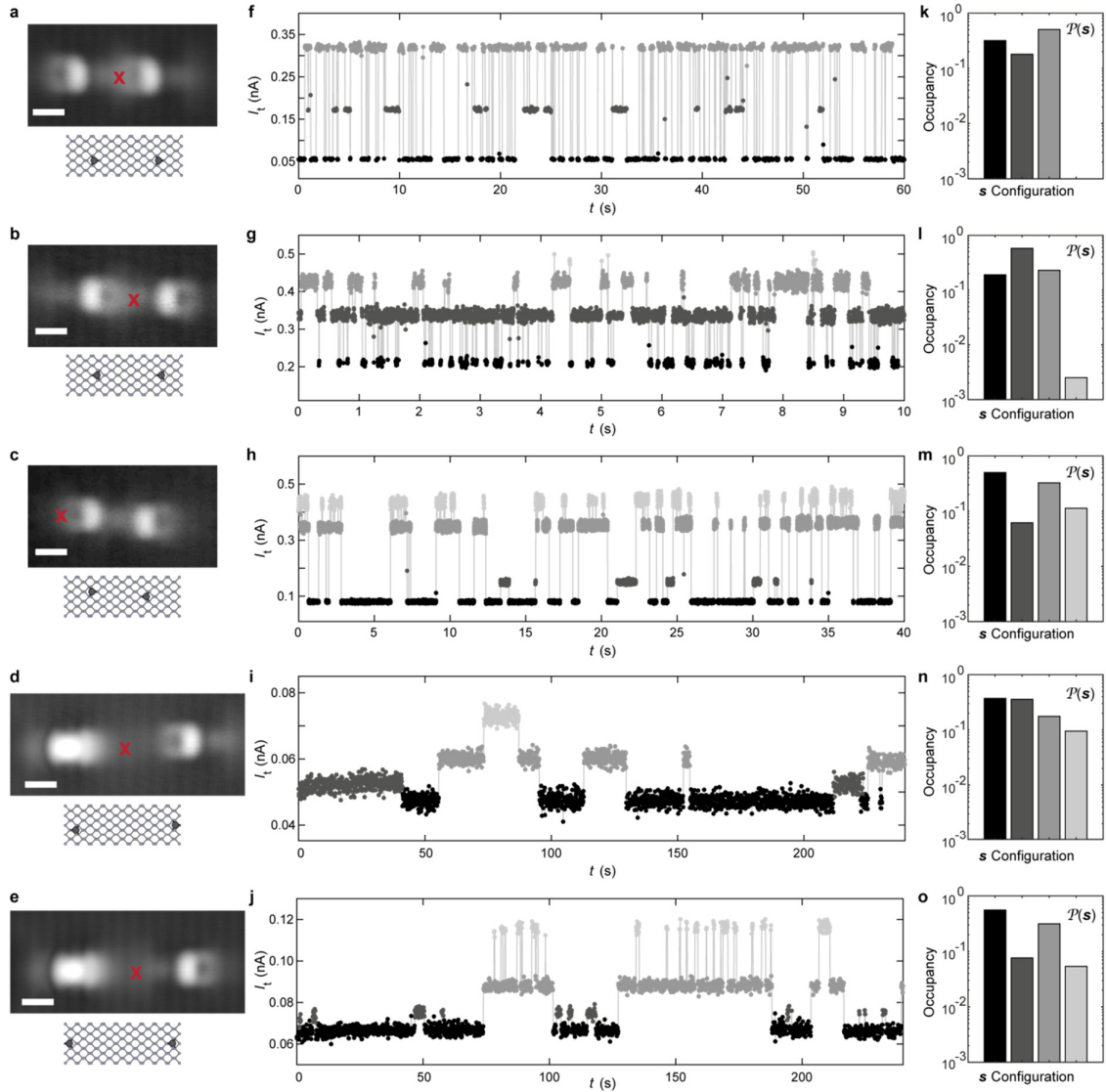

**Supplementary Figure 4. Neural dynamics of various interatomic separations between two Co atoms on BP.** Panels **a-e** show STM topographies of several configurations of two coupled Co atoms on BP ($V_s$ = -400 mV, scale bar = 1 nm) and models of the atom positions on the BP. **f-j,** $I_t(t)$ measured at the red X in **a-e**. Bias voltage for **f:** $V_s$ = 540 mV, **g, h:** $V_s$ = 550 mV, **i, j:** $V_s$ = 450 mV. **k-o,** Time-integrated probability distribution $\mathcal{P}(s)$. The data in this figure was collected with different microtips.

As reported for a single Co atom[3], the probability distribution $\mathcal{P}(s)$ for a given atomic configuration can also vary depending on the location of the tip. This is seen by comparing the configuration in Supplementary Fig. 5b to the configuration in Fig. 1d, where the same configuration is measured with the same microtip at different tip positions, leading to a different distribution $\mathcal{P}(s)$.

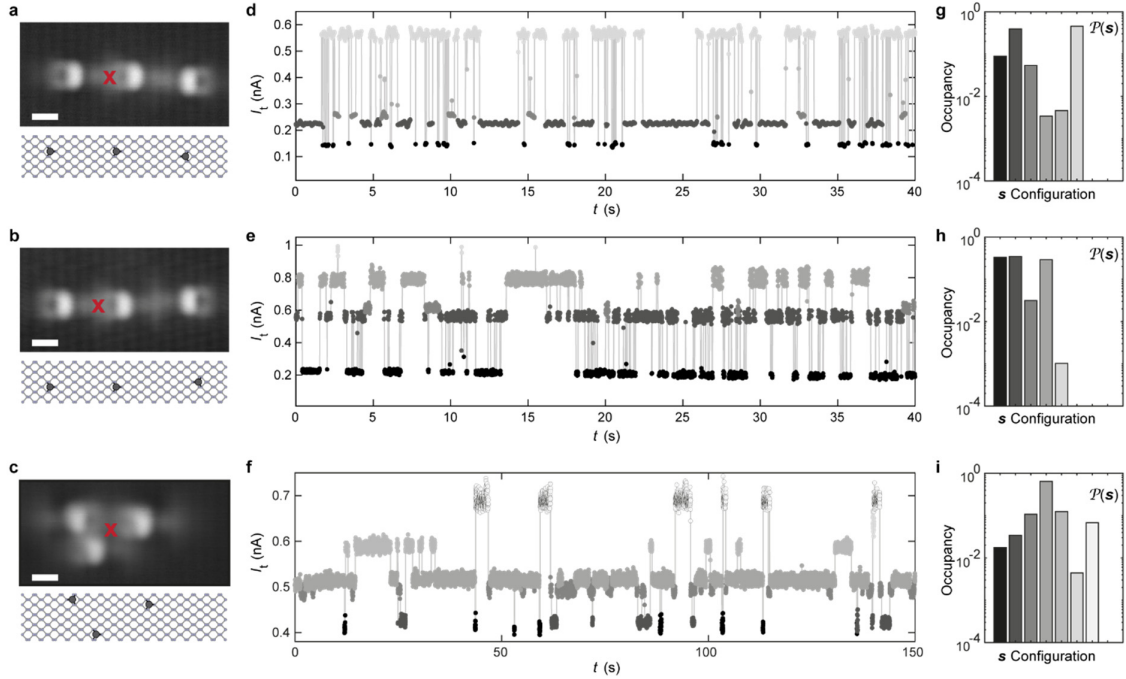

**Supplementary Figure 5. Neural dynamics of various configurations of three coupled Co atoms on BP.**
Panels **a-c** show STM topographies of several configurations of three coupled Co atoms on BP ($V_s$ = -400 mV, scale bar = 1 nm) and models of the atom positions on the BP. **d-f,** $I_t(t)$ measured at the red X in **a-e**. Bias voltage for **d:** $V_s$ = 600 mV, **e:** $V_s$ = 550 mV, **f:** $V_s$ = 450 mV. **g-i,** Time-integrated probability distribution $\mathcal{P}(s)$. The data in this figure was collected with different microtips.

Response to DC stimuli

To characterize the response of single and coupled Co atoms as a function of $V_s$, the probability distribution $\mathcal{P}(s)$ was measured as a function of $V_s$ for various configurations (Supplementary Fig. 6). From the $I_t(t)$ traces, similar to what is seen for individual Co atoms, for higher DC voltages ($V_s$) the switching rate is higher. Furthermore, when increasing the number of coupled atoms, the switching rate decreases (note that the time scales in the $I_t(t)$ traces shown here, are different for the different atomic configurations). As seen in the probability distributions $\mathcal{P}(s)$ for the single atom (Supplementary Fig. 6d, g, j, and m), the state favourability is strongly modified by the applied DC voltage ($V_s$). For the coupled dimer and trimer, the steady-state distribution $\mathcal{P}(s)$ evolves in a complex, non-linear manner with increasing $V_s$. This data indicates that the neural distributions evolve in response to the environment (i.e. the electrical stimuli), thus the probability distributions are conditioned by the environment $\mathcal{P}(s|\varepsilon)$.

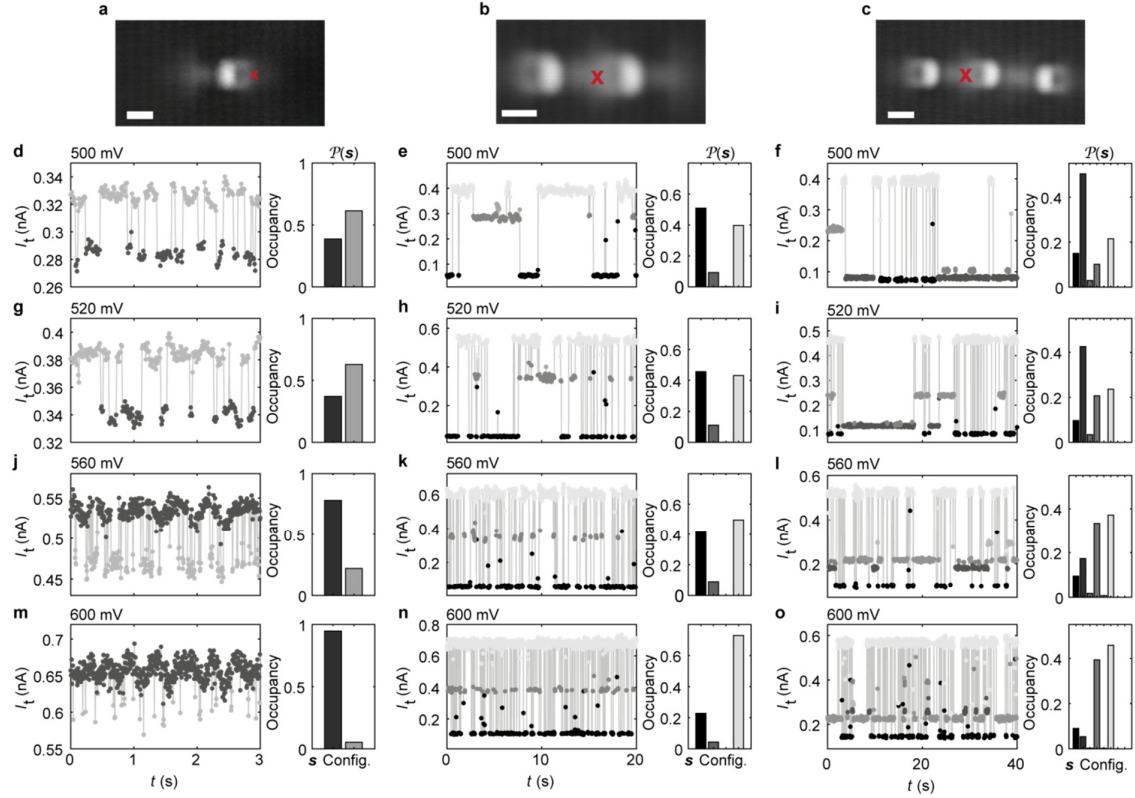

**Supplementary Figure 6. Response of coupled cobalt atoms on BP to varying DC voltages. a-c,** STM topographies of a single Co atom on BP, two coupled atoms and three coupled atoms, respectively ($V_s$ = -400 mV, scale bar = 1 nm). **d-o,** $I_t(t)$ traces (left) and time-integrated probability distributions $\mathcal{P}(s)$ (right) at different DC biases: **d-f** $V_s$ = 500 mV, **g-i** $V_s$ = 520 mV, **j-l** $V_s$ = 560 mV and **m-o** $V_s$ = 600 mV. The $I_t(t)$ traces were measured at the red X's in **a-c**, all with the same microtip. To determine the time-integrated probability distributions, at least 300 switching events were measured. **d, g, j, m** correspond to the configuration in **a**; **e, h, k, n** correspond to the configuration in **b** and **f, i, l, o** correspond to the configuration in **c**. All the data in this figure was collected with the same tip apex.

Influence of the microtip on neural dynamics

In order to understand the influence of the STM tip on the $I_t(t)$ measurement and subsequently the $\mathcal{P}(s|k)$ distributions, we studied the same atomic ensemble from Fig. 2 with a different microtip (Supplementary Fig. 7). We attribute the variations in $\mathcal{P}(s|k)$ to small changes to the local band bending profile (schematically illustrated in Fig. 1). All the data in Supplementary Fig. 7 was collected with the same tip apex. As seen in the modified $w_i$ and $b_i$ of the BM, the modifications to the tip apex and corresponding changes to the local band bending do quantitatively modify the BM. However, the tunable multi-well landscape, seen in the $\mathcal{P}(s|k)$ distributions is robust to these tip changes.

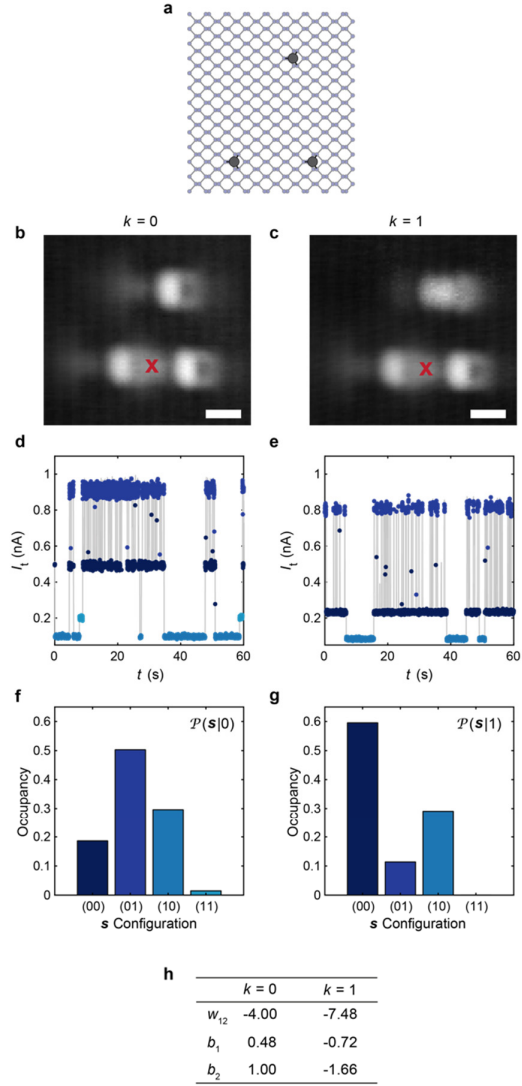

**Supplementary Figure 7. Influence of the microtip on the binary atomic synapse from Figure 2. a,** Schematic of the BP lattice structure with the atomic positions of the coupled Co atoms indicated. **b-c,** Constant-current STM topography with $k=0$ and $k=1$, respectively ($V_s = -400$ mV, scale bar = 1 nm). **d-g,** $I_t(t)$ traces and time-integrated probability distributions $\mathcal{P}(s|k)$ measured with a slightly different tip apex compared to the data shown in Figure 2. **h,** Weights and biases for the two $k$ configurations.

In order to further confirm that the specific termination of the STM tip was not responsible for the self-adaption demonstrated in Fig. 4, we studied the response $\mathcal{P}(k|\varepsilon)$ for the exact same atomic ensemble in Fig. 4 with a different microtip (Supplementary Fig. 8). Again, interpreting the modification of the tip apex as a shift in the local band bending, it is clear that the qualitative change in $\mathcal{P}(\boldsymbol{k}|\varepsilon)$ with $\varepsilon$ observed in Fig. 4 is well reproduced with a different microtip. In this case, the applied offset biases might be slightly shifted with respect to the data in Fig. 4, consistent with a slight offset in the tip's work function.

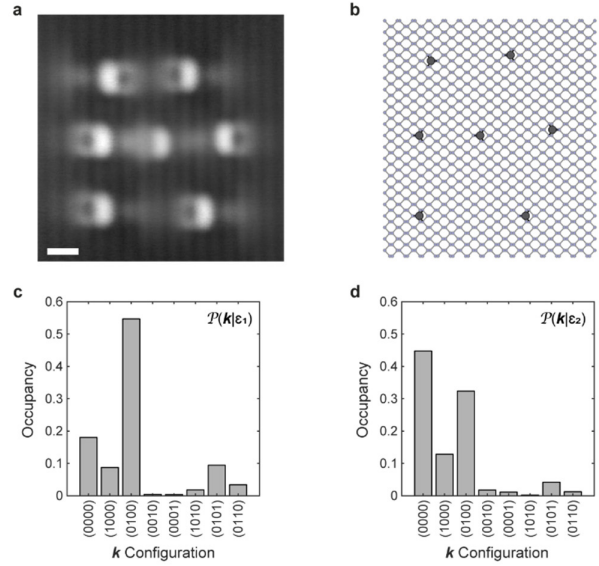

**Supplementary Figure 8. Influence of the microtip on P($\boldsymbol{k}|\varepsilon$) in the Co atom ensemble in Figure 4. a,** Constant-current STM topography of the seven Co atom ensemble in Figure 4 ($V_s$ = -400 mV). **b,** Schematic of the BP lattice structure with the atomic positions of the seven Co atoms indicated. **c-d,** $s$-Integrated probability distribution $\mathcal{P}(\boldsymbol{k}|\varepsilon)$ for $\varepsilon_1 \equiv V_{\text{off}} = 180$ mV, $\varepsilon_2 \equiv V_{\text{off}} = 200$ mV for a slightly different microtip compared to the distributions shown in Figure 4.

## Uncertainty in the probability distributions

The error bars shown in the main text are derived as follows. The entire measurement duration for a single probability distribution was subdivided into $n$ sections; the size of the sections was determined according to the criterion based on the Kullback-Leibler (*KL*) divergence:

$$\frac{1}{n}\sum_{i=1}^{n} KL(p_i(\boldsymbol{s}), p_\infty(\boldsymbol{s})) < 0.05,$$

where $\mathcal{P}_i(\boldsymbol{s})$ is the individual section distribution and $\mathcal{P}_\infty(\boldsymbol{s})$ is the cumulative measurement distribution. The criterion was used to approximate the minimum amount of time needed to measure $I_t(t)$ in order to converge to the steady-state distribution. The final $\mathcal{P}(\boldsymbol{s})$ distribution was derived from the mean value of all $n$ $\mathcal{P}_i(s)$ distributions. For Fig. 1 and Fig. 2, where $n > 10$, the error bars were derived from the 95% confidence interval for a normal Gaussian distribution. As $n < 10$ for Fig. 3 and Fig. 4, we use the 80% and 95% confidence interval defined for a t-distribution with $n$-1 degrees of freedom, respectively. For nearly all probability distributions shown in this work, the data were collected for a minimum of $n$=4 sections, each representing an approximate steady-state distribution. In all experiments, we have taken great care to minimize sources of experimental error related to the tip position and conditioning. In order to obtain an approximation for these sources of uncertainty, we examine the $\mathcal{P}(s)$ distribution for a single atom measured with $n$ = 144 individual measurements (Supplementary Fig. 9). The size of this measurement was designed to minimize the impacts of finite time approximations. In these measurements, the standard error is 0.005. This measurement provides an upper bound of the cumulative effects of these sources of experimental error.

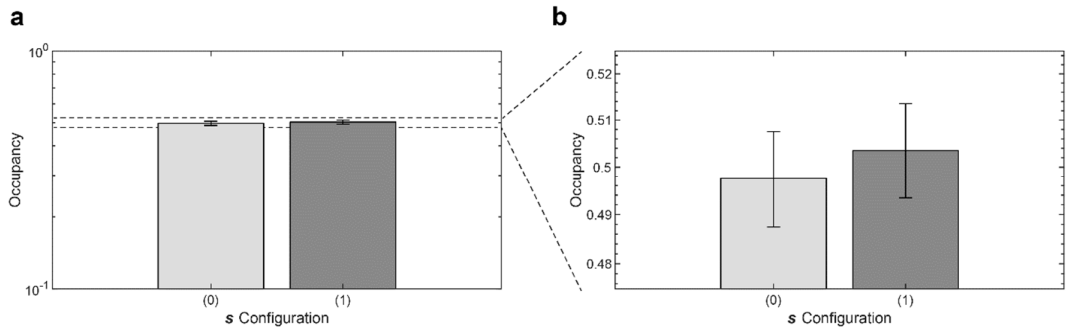

**Supplementary Figure 9. Uncertainty characterization in a single atom. a,** Probability distribution for a single atom with nearly 8,000 switches. The error bars show the 95% confidence interval for 144 separate measurements which reached steady-state conditions according to the criterion described above. **b,** Expanded view of the errors to show the minimum experimental uncertainty in this work. The corresponding standard error is approximately 0.005.

## Time evolution of the probability distribution $\mathcal{P}(k|\varepsilon_3)$

To quantitatively understand the amount of time or number of switching events necessary to measure the steady-state distribution $\mathcal{P}(k|\varepsilon)$, we consider the evolution of $\mathcal{P}(k|\varepsilon_3)$ at selected time intervals from Fig. 4 (Supplementary Fig. 10). After 13% of the total measurement time, the two most favourable states are already clearly identified. After 54% of the time the distribution exhibits only minimal changes.

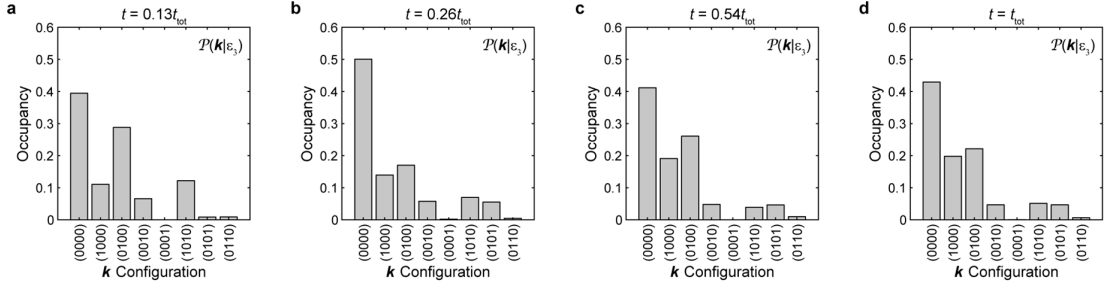

**Supplementary Figure 10. Time-evolution of the probability distribution P(k|ε₃) in Figure 4. a,** $\mathcal{P}(k|\varepsilon_3)$ after 92 minutes in which 61 *k*-switching events occurred (13% of the total measurement time). **b,** $\mathcal{P}(k|\varepsilon_3)$ after 180 minutes in which 101 *k*-switching events occurred (26% of the total measurement time). **c,** $\mathcal{P}(k|\varepsilon_3)$ after 372 minutes in which 156 *k*-switching events occurred (54% of the total measurement time). **b,** $\mathcal{P}(k|\varepsilon_3)$ after 684 minutes in which 340 *k*-switching events occurred (100% of the total measurement time). The data in this figure was collected with the same tip as in Figure 4.

## Boltzmann machine modelling

The Boltzmann machine is a stochastic neural network of binary neurons. Each neuron $i$ takes binary values $s_i = \pm 1$ that denote firing and non-firing. The neural dynamics is given by sequential Glauber dynamics. At each time step, a neuron $i$ is selected at random and its state is set to the new value $s_i'$ with probability $p(s_i'|s) = \sigma(s_i' h_i(s))$ with $s = (s_1,\ldots,s_n)$ the current state of the network.

$$h_i(s) = \sum_{j=1}^{n} w_{ij} s_j + b_i$$

is the summed input of all other neurons on neuron $i$ with $w_{ij}$ the synaptic coupling from neuron $j$ to neuron $i$ and $-b_i$ is a threshold value. $\sigma$ is the sigmoid non-linearity $\sigma(x) = (1 + e^{-2x})^{-1}$.

For finite values of the couplings and thresholds, the Glauber dynamics defines an ergodic Markov process that has a unique stationary distribution[4]. When the coupling matrix $w$ is symmetric ($w_{ij} = w_{ji}$) the Glauber dynamics satisfies detailed balance and the stationary distribution is the Boltzmann distribution

$$p(s|w,b) = \frac{e^{-E(s|w,b)}}{Z(w,b)} \qquad (1)$$

where the energy of the system is

$$E(s|\mathbf{w}, \mathbf{b}) = -\sum_{i>j} w_{ij} s_i s_j - \sum_i b_i s_i$$

and the partition function can be defined as

$$Z(\mathbf{w}, \mathbf{b}) = \sum_s e^{-E(s|\mathbf{w},\mathbf{b})}.$$

Learning the Boltzmann machine is defined as a procedure to find the parameters $\mathbf{w}$, $\mathbf{b}$ such that the Boltzmann distribution Eq. 1 is as close as possible to a given distribution $q$. $q$ is typically given in terms of a data set of $N$ samples $s^\mu$, $\mu = 1, \ldots N$, where each sample $s^\mu$ is a binary vector of length $n$. The data set defines the empirical probability distribution

$$q(s) = \frac{1}{N} \sum_{\mu=1}^N \delta_{s,s^\mu}.$$

In the simplest case with no hidden units, the 'distance' between $p$ and $q$ is defined as the relative entropy or Kullback-Leibler divergence

$$KL(q, p) = \sum_s q(s) \log \frac{q(s)}{p(s)}.$$

Minimizing $KL(q, p)$ with respect to $w$, $b$ is equivalent to maximizing the log likelihood

$$L = \sum_s q(s) \log p(s).$$

The maximization can be performed by gradient ascent on $L$:

$$\Delta w_{ij} \propto \frac{\partial L}{\partial w_{ij}} = \langle s_i s_j \rangle_q - \langle s_i s_j \rangle_p$$

$$\Delta b_i \propto \frac{\partial L}{\partial b_i} = \langle s_i \rangle_q - \langle s_i \rangle_p$$

where $\langle \ldots \rangle_{p,q}$ is the expectation with respect to the Boltzmann distribution $p$ and the empirical distribution $q$, respectively. Learning stops when the gradients are zero, i.e. when the statistics under $p$ and $q$ are equal. This is the well-known Boltzmann Machine learning rule[5].

The learning procedure can be generalized to include hidden units. In this case the state vector $s = (v, h)$ with $v$ a vector of visible units and $h$ a vector of hidden units. The learning problem is now to find the marginal distribution

$$p(v) = \sum_h p(v, h)$$

that is closest to the empirical distribution $q(v)$ that is defined on the visible units only.

The Boltzmann machine is an essential computational part of many stochastic neural network models, such as unsupervised learning[6,7] and deep learning[8]. A special case of the BM is the so-called Hopfield model, or attractor neural network[9]. In this case the couplings are not learned by maximizing the likelihood, but are given directly in terms of 'patterns' $\xi^\mu$ as

$$w_{ij} = \frac{1}{N} \sum_\mu \xi_i^\mu \xi_j^\mu$$

These patterns become local minima of the energy $E$ and the Glauber dynamics attracts the network state $s$ towards one of the $\xi^\mu$.

Impact of the environment on the weights and biases of the BM

Using the BM model (described above), we modelled the probability distributions $\mathcal{P}(s|k,\varepsilon_i)$ from Fig. 4d-f to extract the corresponding weights and biases for each individual BM (Supplementary Tables 1-3), defined both by the $k$-configuration and the environment ($\varepsilon$). The data shows that for nearly all $k$-configurations, the weights and biases are significantly different, indicating that the $k$-configuration plays a major role in determining the spin or neuron coupling. The data also indicates that the influence of the environment is strongly dependent on the particular configuration of $k$. For example, the environment drastically modifies the weights and biases for $k$=(0000), while hardly modifying the weights and biases for $k$=(0101).

|        | $k$=(0000) | $k$=(1000) | $k$=(0100) | $k$=(0010) | $k$=(0001) | $k$=(1010) | $k$=(0101) | $k$=(0110) |
|--------|-----------|-----------|-----------|-----------|-----------|-----------|-----------|-----------|
| $w_{12}$ | -33.84 | -14.34 | 11.69 | -11.68 | 45.18 | -22.50 | 25.97 | -8.54 |
| $w_{13}$ | -33.23 | -2.88 | 14.27 | -0.40 | 34.68 | -12.13 | 29.55 | 3.98 |
| $w_{23}$ | -32.56 | -4.62 | 6.14 | -3.21 | 36.01 | -23.45 | 20.54 | 2.13 |
| $w_{123}$ | 38.51 | 3.73 | -15.87 | 1.75 | -39.96 | 17.96 | -28.09 | -1.86 |
| $b_1$ | 28.83 | 1.30 | -16.91 | -1.69 | -45.28 | 12.18 | -36.08 | -6.34 |
| $b_2$ | 30.68 | 5.53 | -10.96 | 3.65 | -42.93 | 17.91 | -22.93 | 0.21 |
| $b_3$ | 30.23 | 3.29 | -11.73 | 1.61 | -32.99 | 15.46 | -25.21 | -3.42 |

**Supplementary Table 1. Weights and biases for seven atom BM from Figure 4 with $V_{off}$ = 160 mV.**

|        | $k$=(0000) | $k$=(1000) | $k$=(0100) | $k$=(0010) | $k$=(0001) | $k$=(1010) | $k$=(0101) | $k$=(0110) |
|--------|-----------|-----------|-----------|-----------|-----------|-----------|-----------|-----------|
| $w_{12}$ | -10.18 | -14.11 | 24.15 | -14.34 | 46.24 | -20.47 | 26.50 | -5.81 |
| $w_{13}$ | 3.57 | 0.06 | 26.34 | -0.13 | 35.29 | -10.33 | 29.11 | 6.81 |
| $w_{23}$ | 1.15 | -1.98 | 14.68 | -1.21 | 38.07 | -22.49 | 20.52 | 4.20 |
| $w_{123}$ | -2.05 | 2.13 | -24.94 | 2.19 | -40.23 | 16.77 | -27.84 | -3.77 |
| $b_1$ | -5.03 | -0.08 | -34.14 | 0.14 | -46.18 | 9.83 | -36.23 | -9.33 |
| $b_2$ | 1.27 | 4.38 | -19.82 | 4.26 | -44.17 | 16.61 | -23.07 | -1.89 |
| $b_3$ | -2.25 | 0.11 | -20.26 | -0.37 | -34.93 | 14.13 | -25.24 | -5.90 |

**Supplementary Table 2. Weights and biases for seven atom BM from Figure 4 with $V_{off}$ = 180 mV.**

|        | $k$=(0000) | $k$=(1000) | $k$=(0100) | $k$=(0010) | $k$=(0001) | $k$=(1010) | $k$=(0101) | $k$=(0110) |
|--------|-----------|-----------|-----------|-----------|-----------|-----------|-----------|-----------|
| $w_{12}$ | -10.78 | -14.10 | 24.10 | -14.55 | 0.00 | -16.48 | 26.26 | -5.97 |
| $w_{13}$ | 2.66 | -0.22 | 26.16 | -0.27 | 0.00 | -5.98 | 29.65 | 6.85 |
| $w_{23}$ | 1.15 | -1.50 | 14.67 | -1.76 | 0.00 | -21.05 | 20.83 | 2.85 |
| $w_{123}$ | -1.49 | 2.15 | -24.72 | 2.32 | 0.00 | 13.92 | -27.80 | -3.66 |
| $b_1$ | -4.18 | 0.11 | -34.07 | 0.23 | 0.00 | 5.06 | -36.36 | -9.55 |
| $b_2$ | 1.50 | 4.14 | -19.76 | 4.63 | 0.00 | 14.20 | -23.06 | -1.17 |
| $b_3$ | -1.93 | -0.06 | -20.25 | 0.07 | 0.00 | 11.57 | -25.80 | -4.99 |

**Supplementary Table 3. Weights and biases for seven atom BM from Figure 4 with $V_{off}$ = 200 mV.**

## Correlations between $\mathcal{P}(s|k)$ distributions

In order to compare the correlations between the $\mathcal{P}(s|k)$ distributions, we use the *KL* divergence, defined above, to quantify the similarity between any two distributions. The larger the value of the *KL* divergence is, the less similar two distributions are. In Supplementary Fig. 11, we plot the magnitude of $KL(\mathcal{P}(s|k_1),\mathcal{P}(s|k_2))$ with configurations $k_1$ defined on the left axis and $k_2$ on the bottom axis. As seen in this figure, the *KL* divergence between all $\mathcal{P}(s|k)$ distributions is greater than one, with the exception of the $\mathcal{P}(s|\mathbf{1000})$ and $\mathcal{P}(s|\mathbf{1010})$. The *KL* divergence between these two distributions is approximately 0.05.

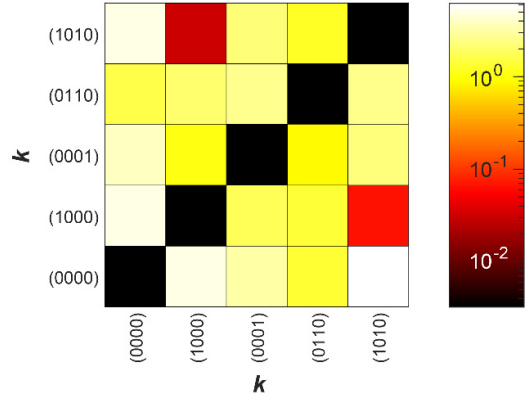

**Supplementary Figure 11.** *KL* **Divergences between** $\mathcal{P}(s|k)$ **Distributions in Figure 3.** *KL* divergence, calculated between two $\mathcal{P}(s|k)$ distributions, with the corresponding $k$ configuration shown on the left and bottom. The distributions from Figure 3 have been smoothed with a Laplace filter before calculating the *KL* divergence.

We additionally studied correlations between the $\mathcal{P}(s|k)$ distributions for the atomic ensemble shown in Fig. 4. Again, analyzing the *KL* divergence between $\mathcal{P}(s|k)$ distributions ($KL(\mathcal{P}(s|k_1),\mathcal{P}(s|k_2))$), we show the results for the three offset biases ($V_{off}$ = 160 mV, $V_{off}$ = 180 mV, $V_{off}$ = 200 mV) studied in Fig. 4. The data shown in Supplementary Fig. 12 clearly shows that the *KL* divergence between any two $\mathcal{P}(s|k)$ distributions is greater than 0.01 for all applied DC biases. Furthermore, there is only a single pair of states, which have *KL* divergences below 0.1 for all three biases. This data indicates that while there are some correlations between $\mathcal{P}(s|k)$ distributions (with *KL* divergence between 0.01 and 0.1), these distributions are also highly sensitive to the electrostatic environment created by the STM tip. This provides a pathway to engineer the system to minimize correlations between the $2^k$ $\mathcal{P}(s)$ distributions.

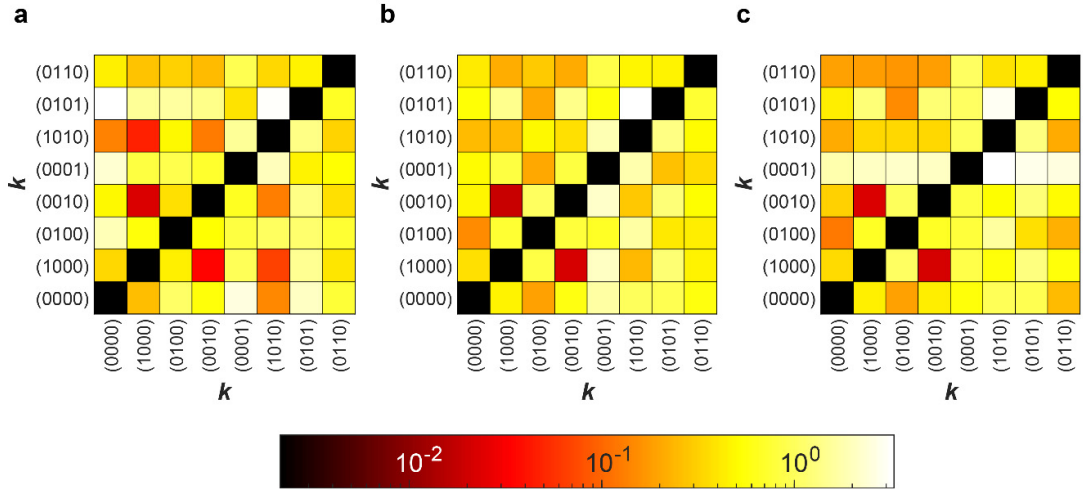

**Supplementary Figure 12. *KL* Divergences between the $\mathcal{P}(s|k)$ Distributions in Figure 4.** *KL* divergences, calculated between two $\mathcal{P}(s|k)$ distributions, with the corresponding *k* configuration shown on the left and bottom for **(a)** $V_{\text{off}}$ = 160 mV, **(b)** $V_{\text{off}}$ = 180 mV, and **(c)** $V_{\text{off}}$ = 200 mV. The distributions from Fig. 3 have been smoothed with a Laplace filter before calculating the *KL* divergence.

## Energy efficiency

The continuous power consumption of the simple seven atom Boltzmann machine shown in Fig. 4 is approximately 50 pW (0.5 V x 0.1 nA). In this system, the operating power is distributed over all the components in the BM, making the power per component on the order of 5 pW. This power can readily be reduced, simply by reducing the currents used to read-out the state of the BM.

While it is difficult to compare the model here to fully implemented neuromorphic devices, we can examine per component power consumption for comparison. The industry standard memristors, used for both neurons and synapses, generally operate in the µW range when used for neurons[10]. While phase change materials, commonly employed for spike-timing dependent approaches to on-chip learning, require power from 0.1-0.5 mW to instigate the phase transition required for neuromorphic functionality[11]. This heavy power burden can be significantly reduced by utilizing ultra-short optical or electrical signals to change the materials phase, resulting in 10's of nJ per synaptic event[12]. Regardless, it is clear that improvements in energy efficiency provide another reason to develop neuromorphic hardware beyond the current hybrid systems.